\documentclass[]{aastex631}
\usepackage{amsmath}

\begin{document}

\title{The heating efficiency of hot Jupiters from a data-driven perspective}

\correspondingauthor{Sheng Jin}
\email{jins@ahnu.edu.cn}

\author[0000-0002-9063-5987]{Sheng Jin}
\affiliation{Department of Physics, Anhui Normal University, Wuhu, Anhui 241002, China}
\affiliation{Purple Mountain Observatory, Chinese Academy of Sciences, Nanjing, Jiangsu 210023, China}

\author[0000-0001-9424-3721]{Dong-Hong Wu}
\affiliation{Department of Physics, Anhui Normal University, Wuhu, Anhui 241002, China}

\author{Yi-Xuan CAO}
\affiliation{High School Affiliated to Nanjing Normal University, Nanjing, Jiangsu 210003, China}

\author{Zi-Yi Guo}
\affiliation{High School Affiliated to Nanjing Normal University, Nanjing, Jiangsu 210003, China}

\begin{abstract}
The inflated radii of hot Jupiters have been explored by various theoretical mechanisms. By connecting planetary thermal evolution models with the observed properties of hot Jupiters using hierarchical Bayesian models, a theoretical parameter called the heating efficiency has been introduced to describe the heating of the interiors of these planets. Previous studies have shown that the marginal distribution of this heating efficiency parameter has a single-peak distribution along the planetary equilibrium temperature ($T_{\rm eq}$). Since the foundation of these Bayesian inference models are the observed properties of hot Jupiters, there must be a corresponding feature in the observed data that leads to the inferred single-peak distribution of the heating efficiency. This study aims to find the underlying cause of the single-peak heating efficiency distribution without relying on specific theoretical models. By analyzing the relationship between different observed physical properties, we obtained a similar single-peak distribution of the radius expansion efficiency of hot Jupiters along $T_{\rm eq}$, which can be explained by the correlation with the stellar effective temperature.
However, a detailed investigation suggests that this single-peak distribution is actually the result of straightforward physical processes.
Specifically, the increase in heating efficiency can be attributed to the increase in incident stellar flux, while the decrease in heating efficiency can be attributed to the rise in gravitational binding energy associated with the increase in planetary mass.

\end{abstract}


\section{Introduction} \label{sec:intro}
Being the first type of exoplanet discovered, hot Jupiters have garnered significant attention due to their unusual orbital and physical properties \citep{Mayor1995,Charbonneau2000,Henry2000}.
Observational data from a variety of detection methods, including spectroscopy,  have helped us to characterize these highly irradiated objects.
Such information can help us to understand how hot Jupiters form and evolve, and to gain a better understanding of Jovian planets in general \citep{Knutson2007,Fortney2008,Fortney2010,Molliere2015,Thorngren2016}.

One puzzling aspect of hot Jupiters is that they have unusually large radii compared to the gaseous planets in our solar system. \citep{Burrows2000,Anderson2011,Enoch2012,Almenara2015,Delrez2016}.
Although the abnormal radii of hot Jupiters can be intuitively explained by the high levels of radiation they receive from their host stars, theoretical models that take the incoming stellar flux into account still predict radii that are smaller than observed \citep{Fortney2007,Baraffe2008,Demory2011,Miller2011,Enoch2012,Weiss2013}. 
A number of mechanisms have been proposed to explain the radius anomaly of hot Jupiters. %
Some require additional energy to heat up the interior of hot Jupiters, which can increase their entropy and thus their radii \citep{Arras2006,Marleau2014}.
For example, the power of tidal dissipation of an eccentric orbit \citep{Bodenheimer2001,Arras2010,Jermyn2017}, 
advection of potential temperature due to strong stellar irradiation \citep{Youdin2010,Tremblin2017,Sainsbury2019},
atmospheric circulation that creates the thermal dissipation of kinetic energy \citep{Guillot2002,Showman2002}, 
Ohmic dissipation of the irradiation energy through a planet's magnetic fields \citep{Batygin2010,Batygin2011,Perna2010,Huang2012,Wu2013,Ginzburg2016}, etc.
Some mechanisms slow down the normal cooling of a planet.
For example, the effect of enhanced atmospheric opacities \citep{Burrows2007},
double diffusive convection \citep{Chabrier2007,Kurokawa2015},
the mechanical greenhouse \citep{Youdin2010}, etc.

Because hundreds of hot Jupiters have been discovered, statistical methods can be used to study the mechanisms that are responsible for their large radii.
Hierarchical Bayesian models are particular helpful in this field as they can separate the natural variation of a physical quantity (called inherent scatter in this paper) from the observational biases that affect the measurement of that quantity (called observational bias) \citep{Kelly2007,Rogers2015,Wolfgang2016}.
Using a hierarchical Bayesian model to analyze the statistical distribution of planetary radii, \citet{Sestovic2018} showed that at a specific irradiation flux, the planetary mass plays a critical role in regulating the magnitude of the planet's radius enlargement.

Departing from purely data-driven statistical models, \citet{Thorngren2018} and \citet{Sarkis2021} integrated the thermal evolution of hot Jupiters into their hierarchical Bayesian model.
They showed that the marginal distribution of the heating efficiency, a parameter that describes how efficiently a planet's interior is heated,  exhibits a peaked shape with respect to the planetary $T_{\mathrm{eq}}$.
The heating efficiency increases as $T_{\mathrm{eq}}$ increases up to a maximum at $T_{\mathrm{eq}}$ of 1500-1800 K, and then decreases as $T_{\mathrm{eq}}$ continues to increase. 
While the single-peak distribution of the heating efficiency can be partially explained by various heating mechanisms of the planetary interior, a precise relationship between these mechanisms and the heating efficiency distribution has not yet been established.

When incorporating planetary thermal evolution in hierarchical Bayesian models, a substantial number of hyperparameters are also introduced in the statistical model.
This can obscure the features that are directly related to the observed physical quantities due to the hierarchical dependence of hundreds of parameters. 
Because the single-peak distribution of the heating efficiency is inferred from the hyperparameters that are on top of directly observed physical quantities \citep{Thorngren2018,Sarkis2021}, there must be a corresponding feature in the directly observed quantities that can explain the peaked marginal distribution of the heating efficiency.
In this work, we search through the directly observed physical quantities of hot Jupiters to find the cause of the peaked shape of the marginal distribution of the heating efficiency.
We aim to identify and characterize the factors that determine the radius expansion of hot Jupiters  using a data-driven approach..

This article is structured as follows: Section \ref{sec:model} describes and validates our two-level, two-parameter hierarchical model.
Section \ref{sec:results} demonstrates that the radius expansion of a hot Jupiter is sensitive to both the effective temperature of its host star and the amount of incoming stellar flux the planet receives. Furthermore, the strength of this dependence on stellar effective temperature can reproduce the single-peak distribution of the heating efficiency along the $T_{\textrm{eq}}$ of hot Jupiters.
Section \ref{sec:analyse} analyzes the dependence of relevant physical properties, particularly focusing on stellar effective temperature.
It rectifies that the single-peak distribution observed in our model can be attributed to fundamental physical mechanisms, namely the combined influence of incoming stellar flux and planetary mass.
Section \ref{sec:summary} provides a concise discussion and summary of the work.

\section{Modeling and Validation}
\label{sec:model}

\subsection{Two-Level, Two-Parameter Hierarchical Model}

In the context of observed data related to a specific physical quantity of a population of objects, at least two sources of variation are anticipated.
The first source of variation is the natural dispersion of the physical quantity itself.
For a quantity that can be influenced by numerous factors, such as the radius of a hot Jupiter, it is reasonable to model this variation as a normal distribution, following the central limit theorem.
The second source of variation is observational bias, which can also be modeled as a normal distribution using maximum likelihood estimation.

Hierarchical Bayesian models are effective at distinguishing between the intrinsic variability of a physical quantity and its observational bias. As a result, they are commonly used to accurately derive the radius relationships of exoplanets \citep{Rogers2015, Wolfgang2016, Chen2017, Sestovic2018}.

To effectively distinguish the inherent scattering of planetary radii from observational bias, we utilize a straightforward two-level, two-parameter hierarchical Bayesian model to examine how different physical factors affect the radii of hot Jupiters. 
This two-parameter approach allows us to focus on the relationship between planetary radius and one specific physical variable at a time. 
This method effectively marginalizes the influences of all other physical factors related to the variable under investigation.
The model can be outlined as follows:

\begin{equation}
y_{\text{obs}} \sim\mathcal{N}(y, ~~ \sigma_{y_{\text{obs}}}) \\
\label{eq:1} 
\end{equation}

\begin{equation}
x_{\text{obs}} \sim\mathcal{N}(x, ~~ \sigma_{x_{\text{obs}}})
\label{eq:2} 
\end{equation}

\begin{equation}
y\sim\mathcal{N}(C x^{\gamma}, ~~ \sigma_{y})	  \label{eq:3} 
\end{equation}

The first level of the hierarchical model  is defined by Equations \ref{eq:1} and \ref{eq:2}. This level  includes two variables, $x$ and $y$, each associated with observational standard deviations $\sigma_{x_{\text{obs}}}$ and $\sigma_{y_{\text{obs}}}$, respectively. 
These two variables have been recorded as the observed sets $x_{\text{obs}}$ and $y_{\text{obs}}$. 
The second level of the hierarchical model is described by Equation \ref{eq:3}, which illustrates a power-law relationship between the variables $x$ and $y$.
We employ a power-law function to model the relationship between various pairs of physical quantities, as it effectively captures a broad range of dependencies with notable precision.
Additionally, its concise formulation makes it well-suited for examining the radius relationships of hot Jupiters \citep{Laughlin2011}.

When applying this model to analyze the radius relationships of hot Jupiters, $y$ represents the mean planetary radius corresponding to a specific value of the physical quantity $x$.
The variable $x$ can encompass a variety of physical quantities, such as irradiation flux, planetary mass, stellar effective temperature, and others.
It is important to note that, in reality, the Gaussian errors should be modeled independently for each planet and each observation.
The hierarchical model outlined in Equations \ref{eq:1}-\ref{eq:3} simplifies the planetary radius relationship by assuming that the radii of the entire hot Jupiter population can be approximated as a single Gaussian distribution.
This simplification allows for effective inference of the general characteristics of hot Jupiter populations \citep{Sestovic2018,Thorngren2018,Sarkis2021}.

Based on the aforementioned two-level, two-parameter model, the corresponding hierarchical inference model is as follows:
                    
\begin{align}
	P(y_{obs}\, | \, y, \, \sigma_{y_{obs}} )
			&\sim \mathcal{N}(y, \, \sigma_{y_{obs}}) \label{eq:sm1} \\
   	P(y\, | \, x, \, C, \, \gamma, \, \sigma_{y} )
			&\sim \mathcal{N}(Cx^{\gamma}, \, \sigma_{y}) \label{eq:sm2} \\
   	P(x\, | \, x_{obs}, \, \sigma_{x_{obs}} )
			&\sim \mathcal{N}(x_{obs}, \, \sigma_{x_{obs}}) \label{eq:sm3} \\
   	P(\gamma) &\sim \mathcal{N}(1, \, 1) \label{eq:pr1} \\
	P(\ln(C)) &\sim \mathcal{U}(-4, \, 4) \label{eq:pr2} \\
    P(\ln(\sigma_{y_{obs}})) &\sim \mathcal{U}(-4, \, 4) \label{eq:pr3} \\
  P(\ln(\sigma_{y})) &\sim \mathcal{U}(-4, \, 4) \label{eq:pr4} \\
   P(\ln(\sigma_{x_{obs}})) &\sim \mathcal{U}(-4, \, 4) \label{eq:pr5} \\
   P(y_{obs}\, | \, \sigma_{y_{obs}}, \, C, \, \gamma, \, \sigma_{y}, \, \sigma_{x_{obs}}, \, x_{obs} ) 
            & \propto P(y_{obs}\, | \, y, \, \sigma_{y_{obs}} ) \times P(y\, | \, x, \, C, \, \gamma, \, \sigma_{y} ) \times P(x\, | \, x_{obs}, \, \sigma_{x_{obs}} )  \nonumber &\\
            & \, \quad \times P(\gamma) \times P(\ln(C)) \times P(\ln(\sigma_{y_{obs}})) \times P(\ln(\sigma_{y})) \times P(\ln(\sigma_{x_{obs}})) \label{eq:post}
\end{align}

where Equation \ref{eq:post} provides the posterior distribution of the model.
The subsequent two subsections will illustrate the retrieval capabilities of this hierarchical model, as well as the simplifications employed when using it to investigate the radii of hot Jupiters. 

\subsection{Grid Test of Hierarchical Inference}

   \begin{figure*}
   \centering
   \includegraphics[width=6.5in]{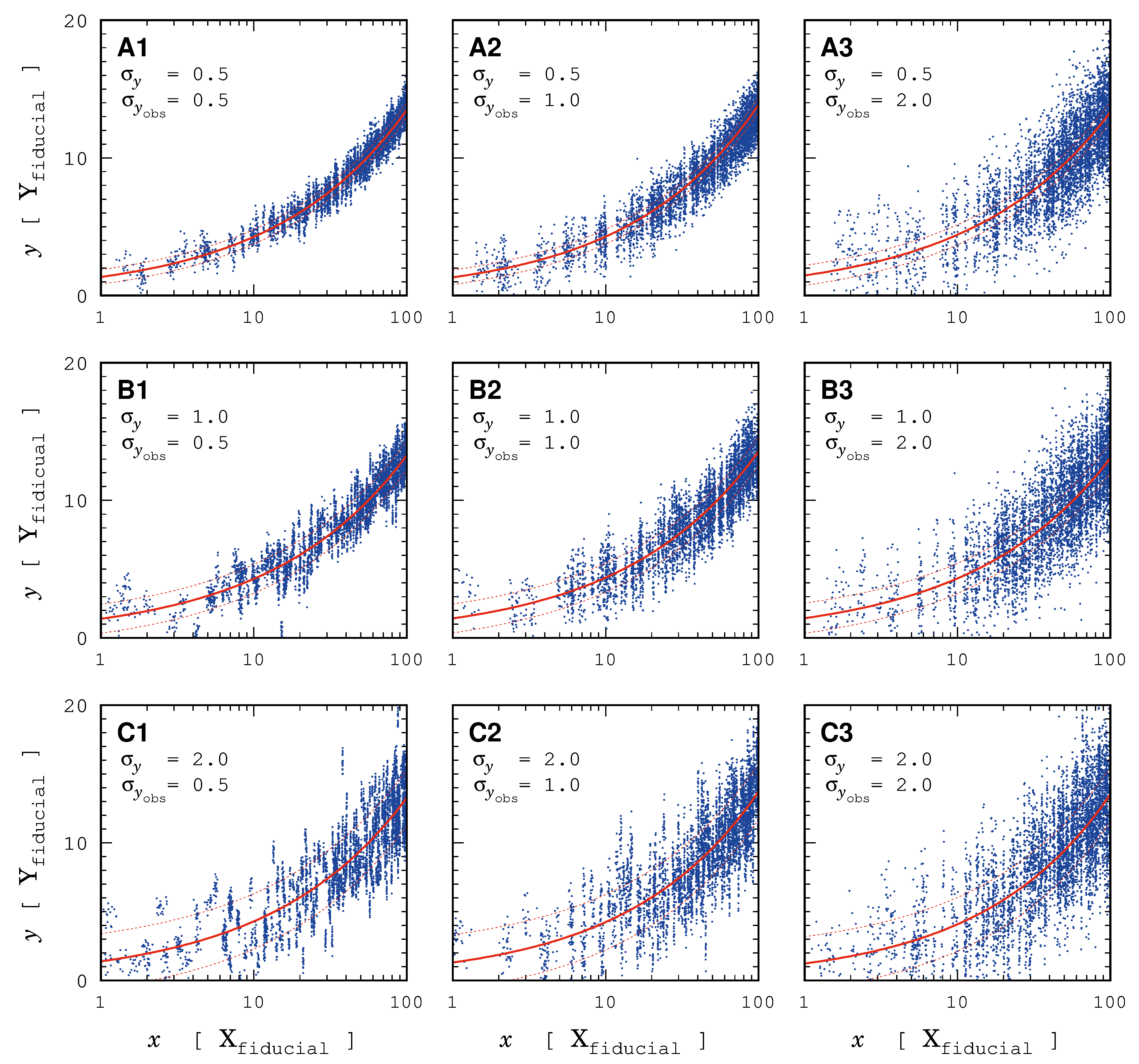}
   \caption{
  A grid test was performed to evaluate the fitting of the two-level, two-parameter hierarchical model using simulated data generated by various combinations of $\sigma_{y}$ and  $\sigma_{y_{\rm{obs}}}$.
  In each panel, the blue dots represent the simulated data points, while the red line shows the power-law relationship given by the fitted values of $log(C)$ and $\gamma$ ($y$, $x$). The red dotted line denotes the one sigma regions of the power-law relationship, computed from the fitted $\sigma_{y}$.
  }
              \label{fig1}%
    \end{figure*}

\begin{table*}
    \caption{Grid test for the effectiveness of our hierarchical inference model in differentiating between intrinsic and observational scatters of $y$.}
\label{table:1}      
\centering          
\begin{tabular}{c c c c c c c c}     
\hline
\hline
    RUN & ${\sigma}_y$(model) & ${{\sigma}_y}_{\mathrm{obs}}$(model) & ${\sigma}_y$(fitted) & ${{\sigma}_y}_{\mathrm{obs}}$(fitted) & ${\gamma}$(fitted) & $log({\mathrm{C}})$(fitted) & ${{\sigma}_x}_{\mathrm{obs}}$(fitted) \\
\hline                    
   A1 & 0.5 & 0.5  & 0.527 & 0.487 & 0.50 & 0.30 & 0.097  \\  
   A2 & 0.5 & 1.0  & 0.533 & 0.970 & 0.51 & 0.28 & 0.098  \\
   A3 & 0.5 & 2.0  & 0.726 & 1.935 & 0.48 & 0.38 & 0.097  \\
   A3+  & 0.5 & 2.0  &  0.577 & 1.974 & 0.50 & 0.28 & 0.096  \\
\hline
   B1 & 1.0 & 0.5  & 1.051 & 0.477 & 0.49 & 0.33 & 0.099  \\
   B2 & 1.0 & 1.0  & 1.051 & 0.961 & 0.49 & 0.35 & 0.097  \\
   B3 & 1.0 & 2.0  & 1.094 & 1.954 & 0.48 & 0.36 & 0.096  \\
   \hline       
   C1 & 2.0 & 0.5  & 2.014 & 0.482 & 0.52 & 0.21 & 0.096  \\
   C2 & 2.0 & 1.0  & 2.014 & 0.970 & 0.51 & 0.27 & 0.097  \\
   C3 & 2.0 & 2.0  & 1.954 & 1.954 & 0.52 & 0.21 & 0.096  \\
\hline                  
\hline                  
\end{tabular}
\end{table*}

One question concerning our hierarchical inference model is whether it can effectively differentiate between the two levels of dispersion.
To investigate this, we conducted a grid test utilizing simulated data generated by Equation \ref{eq:1}-\ref{eq:3}, employing various combinations of  $\sigma_{y}$ and $\sigma_{y_{\rm{obs}}}$.
The values for $log(C)$, $\gamma$ ($y$, $x$), and $\sigma_{x_{\rm{obs}}}$ for all test runs were fixed at 0.3, 0.5, and 0.1, respectively.
For each value of $x$, an average of 18 simulated observations of $y$ were generated.
The simulated data for the grid test is presented in Figure \ref{fig1}.
For each run, we use the parallel tempering MCMC code Nii-C \citep{Jin2024} to sample the posterior distributions specified by Equation \ref{eq:post}.
In total, 5,000,000 iterations were sampled for each run, with the first 1,000,000 were discarded as intial burn-in.
We examined the corner plot of each run to verify that convergence was achieved. 
A typical corener plot illustrating the mass-radius relationship fitting is displayed in Figure \ref{fig2},  while similar corner plots from other runs have been omitted for simplicity.
In section \ref{sepbin}, we also computed the Gelman-Rubin criterion for the separate binning group (Table \ref{tab:RC}), which further confirms that our parallel tempering MCMC have reached convergence.
We summarize the fitted mean values of each parameter in Table \ref{table:1}. 
The power-law relationship, along with the one sigma regions of this relationship  derived from the fitted values of $log(C)$, $\gamma$ ($y$, $x$), and $\sigma_{y}$,  is presented in Figure \ref{fig1}.

We find that the ability to differentiate between the two levels of dispersion depends on the relative magnitudes of the intrinsic scatter and observational bias of $y$. 
This is demonstrated by the fitted values of $\sigma_{y}$ and $\sigma_{y_{\rm{obs}}}$ for each run, as shown in Table \ref{table:1}.
The only run that did not yield accurate values for $\sigma_{y}$ and $\sigma_{y_{\rm{obs}}}$ is A3.
In this run, the simulated data was generated with a $\sigma_{y}$ of 0.5 and a $\sigma_{y_{\rm{obs}}}$ of 2.0, indicating a significant observational bias coupled with a small intrinsic scatter.
However, the fitting result can be significantly improved by increasing the number of observations.
In an additional A3+ run, we generated an average of 48 simulated observations of $y$ for each value of $x$.
In this scenario, the fitted values of $\sigma_{y}$ and  $\sigma_{y_{\rm{obs}}}$ were 0.577 and 1.974, respectively, which are in close proximity to the original model.

When we apply this two-level, two-parameter hierarchical model to the hot Jupiter population, $\sigma_{y}$ and  $\sigma_{y_{\rm{obs}}}$ represent the standard deviations of the mean and observed planetary radii ($\sigma_{R}$ and $\sigma_{R_{obs}}$) for a given value of an independent physical quantity.
We find that the fitted values of $\sigma_{R}$ and $\sigma_{R_{obs}}$ depend on the specific independent physical quantity, such as the incoming stellar flux, planetary mass, stellar effective temperature, and others.
In all instances, the value of  $\sigma_{R}$ is 3 to 5  times greater than that of $\sigma_{R_{obs}}$, suggesting that our two-level, two-parameter model is appropriate for retrieving the intrinsic radii of hot Jupiters.

\subsection{Marginalisation over Planetary Mass}
\label{pmass}

   \begin{figure}
   \centering
   \includegraphics[width=7.10in]{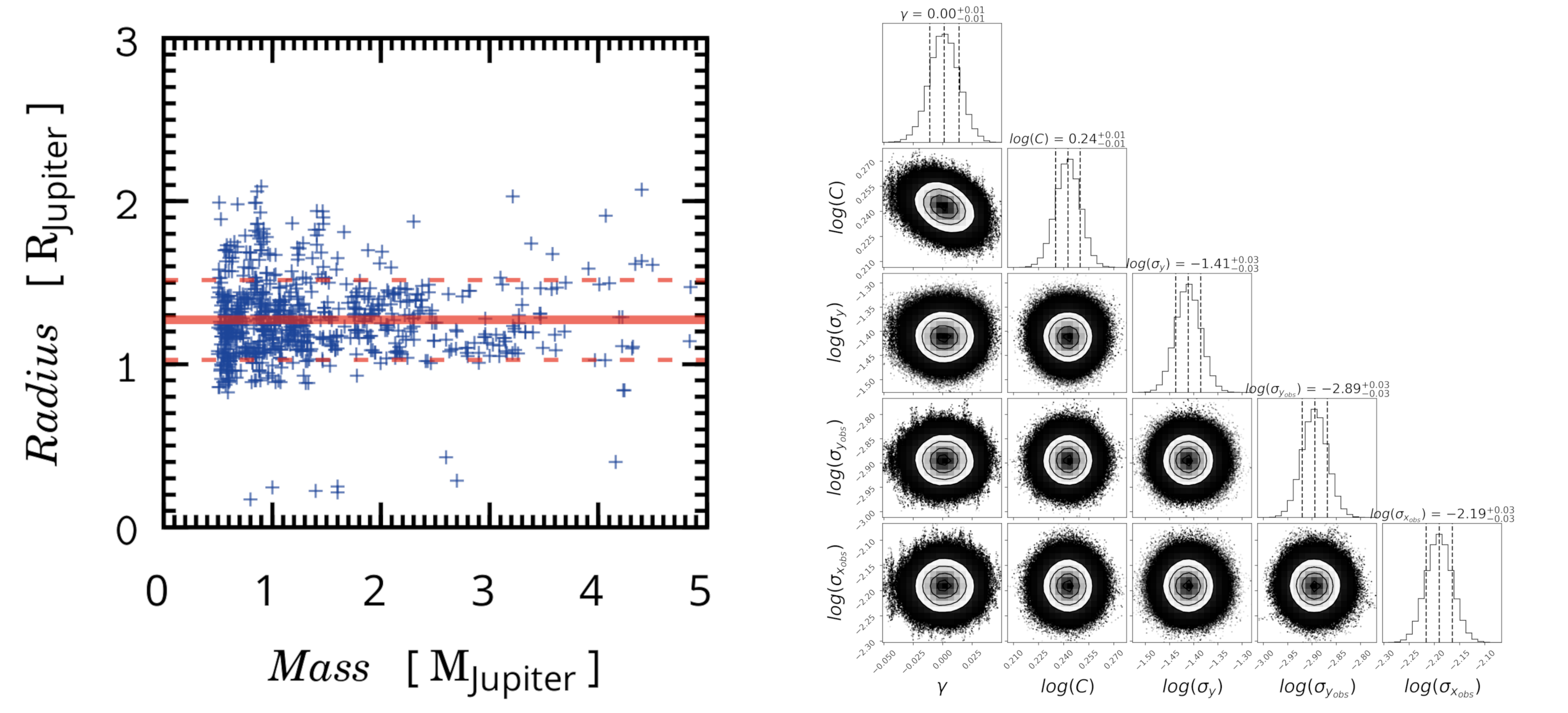}
   \caption{
The left panel displays the mass-radius distribution of hot Jupiters, with mass ranging from 0.5 to 5.0 $M_{\rm Jupiter}$, along with the power-law relationship fitted using the two-level, two parameter hierarchical models.
 Blue dots represent the hot Jupiters. 
 The red line indicate the fitted power-law relationship, and the red dashed lines depict the one sigma regions of this relationship.
 The fitted power-law index, $\gamma$  (${R_{\rm planet}}$, ${M_{\rm planet}}$), is precisely equal to  0.00$^{+0.01}_{-0.01}$. 
 The right panel shows the corner plot derived from parallel tempering MCMC sampling.
}
              \label{fig2}%
   \end{figure}

The fundamental assumption of our two-level, two-parameter hierarchical Bayesian inference model is that the effects of all other physical quantities can be disregarded when analysing the relationship between the radii of the hot Jupiter population and a specific physical quantity.
This is indeed a bold assumption.
Previous work that investigate the radius relationship of hot Jupiters from a data-driven perspective have utilized both planetary mass and incoming flux as independent variables \citep{Sestovic2018}.
In the work of \citet{Sestovic2018}, the authors primarily concentrate on the impact of incoming stellar flux, and as a result, they fit the radius relationships of hot Jupiters separately across different mass ranges (mass bin).
In this study, we utilize all the hot Jupiters with masses between 0.5 and 5 $M_{\rm Jupiter}$ when analysing the radius relationship for a given physical quantity, rather than binning them into different mass ranges.
Our approach assumes that the mass of a hot Jupiter has a negligible influence on its radius expansion when evaluated at the population level, a notion that is supported by the distribution of hot Jupiter radii as a function of planetary mass.

Figure \ref{fig2} shows the mass-radius relationship for all hot Jupiters with mass ranging from 0.5 to 5.0 $M_{\rm Jupiter}$.
The prototype function used to fit this relationship is the power-law function given by Equation \ref{eq:3}, which is suitable for fitting a diverse range of functional forms with a relatively high accuracy.
The fitted power-law index, $\gamma$ (${R_{\rm planet}}$, ${M_{\rm planet}}$), is equal to 0.00 with a standard deviation of 0.01 for the mass-radius relationship. This indicates that the mean radius of hot Jupiters is not correlated with planetary mass.
Such a fitted flat relationship reproduces the mass-radius power-index for Jovian worlds identified by \citet{Chen2017}, which reported a value of $-0.04\pm0.02$, showing that the radius of Jupiter-mass exoplanet is nearly degenerate with respect to planetary mass.
The weak correlation between the population-wide mean radius and the planetary masses of hot Jupiters lends support to our two-parameter model.

It is important to note that the weak dependence of planetary radius on mass is statistically valid only when considering the entire population of hot Jupiters collectively.
In fact, the differences in gravitational potential energy associated with varying planetary masses do influence the extent of radius expansion observed in hot Jupiters. 
If we partition the entire hot Jupiter population into smaller groups, as we do in Sections \ref{single-peak1} and \ref{sec:analyse}, we may inadvertently introduce bias into the results. 
This, in turn, could lead to incorrect interpretations if the mass distributions of the various groups are not adequately considered.
In Section \ref{sec:analyse}, we will demonstrate that the increase in planetary mass for hot Jupiters at ultra-high $T_{\rm eq}$ is likely the primary factor contributing to the previously inferred decrease in heating efficiency at higher $T_{\rm eq}$.

\subsection{Data Binning in Incoming Flux}
\label{fluxbin}

In addition to planetary mass, several other factors also affect the radius of a hot Jupiter.
The rationale behind using a simple two-parameter inference model is to circumvent the mutual dependence among different parameters.  
However, caution is necessary when fitting a two-parameter relationship model, as a third parameter may implicitly affect the two-parameter relationship being examined.
Therefore, for a specific physical parameter that is known to influence the radius of a hot Jupiter, it is necessary to group hot Jupiters based on that parameter.
We can then fit two-parameter relationships between other physical parameters for the grouped subpopulations of hot Jupiters that share similar values of the specific physical parameters known to affect their radii. 
One such specific parameter that can influence the radius of a hot Jupiter is the strength of the incoming stellar flux \citep{Thorngren2018,Sestovic2018,Sarkis2021}. 
Consequently, we categorize the hot Jupiters into different groups based on the amount of flux they receive when analyzing the two-parameter relationships between  planetary radius and other physical quantities. We then fit the power-law indices spearately for the hot Jupiters in these different groups.
This specific binning strategy employed in this work can be referred to as flux binning.

To effectively compare the strength of the incoming flux received by hot Jupiters orbiting around different types of stars at varying orbital semi-major axes, we employ a normalised incoming stellar flux $F$, defined as follows:
\begin{equation}
F = \left( \frac{R_{\rm *}}{a} \right)^2 \sigma T_*^4 \label{eq:incident-flux}
\end{equation}
where $a$ is the orbital semi-major axis, $\sigma$ is the Stefan–Boltzmann constant, $T_{\rm *}$ and $R_{\rm *}$ represent the temperature and radius of the host star, respectively.
Additionally, $F$ is standardized to 1 at a distance of 0.1 AU from a solar-like star with an effective temperature of 5777 K, which corresponds to 100 times the flux received by the Earth.

\section{Single-peak distribution: Initial Observation}

\label{sec:results}

\subsection{Dependence on Flux affected by $T_{\rm star}$}
\label{rad_flux_byTstar}

   \begin{figure*}
   \centering
   \includegraphics[width=6.80in]{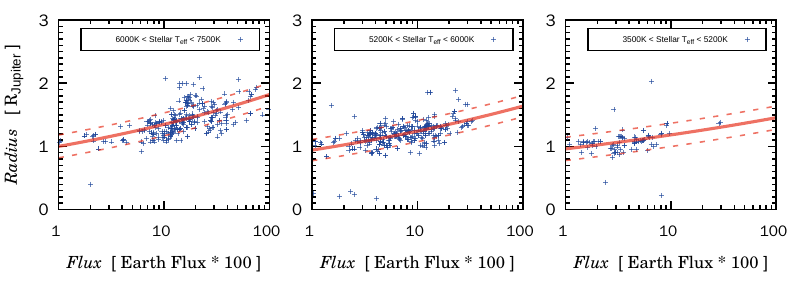}
   \caption{
   The flux-radius distribution of hot Jupiters (blue points) orbiting stars with varying effective temperatures, along with the fitted power-law relationship (red lines) derived from the two-level, two parameter hierarchical models, is displayed.
    The red dashed lines indicate the one sigma regions of the relationship.
     The fitted power-law index, $\gamma$ (${R_{\rm planet}}$, $flux$), for hot Jupiters orbiting host stars with effective temperatures in the ranges of 6500-7500 K, 5200-6000 K, and 3500-5200 K are 0.13$^{+0.01}_{-0.01}$, 0.10$^{+0.01}_{-0.01}$, and 0.08$^{+0.03}_{-0.02}$, respectively.
The incident flux have been normalized to 100 times the amount received by the Earth.
}
              \label{fig3}%
    \end{figure*}

Our normalized incoming stellar flux provides only rough estimation of the total radiation that a hot Jupiter receives, and  it does not capture the variations in the types of radiation experienced by each planet.
Specifically, different types of stars emit different spectral profiles, resulting in the intensity of their radiation being allocated differently across various spectral bands.
In this section, we examine the correlation between the planetary radius and the incoming stellar flux for hot Jupiters orbiting different types of stars. 
Our goal is to determine whether the spectral types of the host stars have a significant impact on the efficiency of radius expansion in hot Jupiters.

Figure \ref{fig3} shows the relationship between planetary radius and the incoming flux for hot Jupiters orbiting different types of stars, as fitted by our two-level, two-parameter hierarchical relationship model.
In this context, $x$ denotes the incoming stellar flux, while $y$ denotes the planetary radius, based on our two-level, two-parameter model.
For host stars with effective temperatures ranging from of 6500 to 7500 K, 5200 to 6000 K, and 3500 to 5200 K, the fitted power-law index $\gamma$  (${R_{\rm planet}}$, $flux$) is 0.13$^{+0.01}_{-0.01}$, 0.10$^{+0.01}_{-0.01}$, and 0.08$^{+0.03}_{-0.02}$, respectively.
It appears that as the stellar effective temperature increases, the correlation between the extent of radius expansion in hot Jupiters and the normalized incoming flux becomes stronger.
At first glance, it seems that the stellar effective temperature also influences the radius expansion of hot Jupiters. 
A more thorough investigation of this topic will be undertaken in the following sections.

\subsection{Dependence on $T_{\rm star}$ affected by Flux}
\label{spectral_teq}

   \begin{figure*}
   \centering
   \includegraphics[width=6.80in]{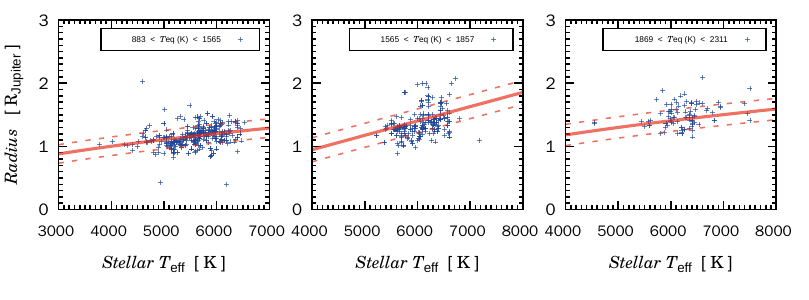}
      \caption{
   The distribution of stellar effective temperature versus planetary radius for hot Jupiters (blue dots) with $T_{\rm eq}$ in the range of 883-1565  K, 1565-1857 K, and 1869-2311 K, respectively. 
   The red lines and red dashed lines show the fitted power-law relationships along with the one sigma regions of those relationships. 
   The fitted power-law index, $\gamma$ (${R_{\rm planet}}$, $T_{\rm star}$), for the three groups of hot Jupiters is 0.45$\pm0.08$, 0.97$\pm0.15$, and 0.43$\pm0.16$, respectively.
              }
         \label{fig4}
   \end{figure*}
   
The results presented in section \ref{rad_flux_byTstar} highlight the inherent difficulties in accurately evaluating how the radii of hot Jupiters depend on specific physical parameters.
Although we have made attempts to establish the relationship between the radii of hot Jupiters and the intensity of incoming stellar radiation, it appears that the effective temperature of the host star also has an indirect impact on this relationship, as shown in Figure \ref{fig3}.
In this section, we explore the influence of stellar radiative properties on the radius of hot Jupiters from a different angle, emphasizing the impact of stellar effective temperature. 
We categorize the hot Jupiters into three groups based on the amount of incoming flux they have received, and individually fit the relationship between planetary radius and stellar effective temperature for each groups, using the flux binning strategy outlined in section \ref{fluxbin}.

The amount of incident flux received by a planet is normalised using the equation \ref{eq:incident-flux}.
The hot Jupiters are categorized into three groups according to the levels of normalized incoming stellar flux they receive.
The first group receives flux values ranging from 1 to 10, the second from 10 to 20, and the third from 20 to 50. 
For each of these groups, we fitted the relationship between the stellar effective temperature and the planetary radius using our two-level, two-parameter hierarchical model, where $x$ represents the stellar effective temperature and $y$ represents the planetary radius in this context.

Since the normalised incoming flux can be converted to the planetary equilibrium temperatures $T_{\rm eq}$, which provide a clearer physical interpretation, we will refer to the three groups with normalised fluxes of 1-10, 20-20, and 20-50 as the subpopulations of  hot Jupiters with $T_{\rm eq}$ in the range of 883-1565 K, 1565-1857 K, and 1869-2311 K, respectively.
Figure \ref{fig4} shows the three groups of hot Jupiters along with the fitted curves representing the relationship between planetary radii and stellar effective temperature.
The mean values of the power-law index $\gamma$ (${R_{\rm planet}}$, $T_{\rm star}$) fitted using our hierarchical power-law relationship are 0.45, 0.97 and 0.43, with corresponding standard deviations of 0.08, 0.15 and 0.16 for each group.
The fitted values of $\gamma$ (${R_{\rm planet}}$, $T_{\rm star}$) for these three groups indicate that it does vary monotonically with the amount of incoming stellar flux. 
In the second group, 
where the planetary $T_{\rm eq}$ ranges from 1565 to 1857 K, the positive correlation between the extent of radius expansion and the stellar effective temperature is the strongest.
In section \ref{rad_flux_byTstar}, Figure \ref{fig3} demonstrates that the radius expansion of hot Jupiters becomes more significant as stellar effective temperatures rise, showing a correlation between radius expansion efficiency and stellar effective temperature.
Here, Figure \ref{fig4} indicates that for hot Jupiters subjected to extremely intense stellar radiation, which leads to ultra-high planetary $T_{\rm eq}$, the correlation between their radius expansion efficiency and stellar spectral type is relatively weak. 
This implies that at extremely high $T_{\rm eq}$, there could be a fundamental physical mechanism that mitigates the effects of radius expansion.

\subsection{Single-Peak Distribution}
\label{single-peak1}

   \begin{figure*}
   \centering
   \includegraphics[width=5.0in]{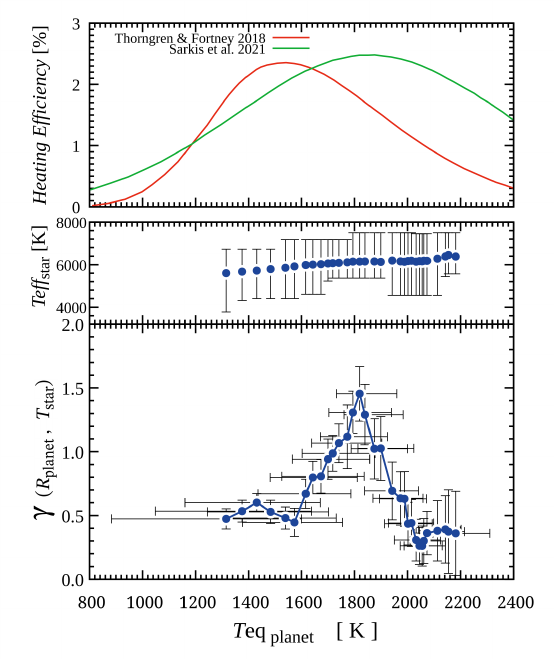}
      \caption{
      For different groups of hot Jupiters with $T_{\rm eq}$ within different ranges, the bottom panel shows the values of $\gamma$ (${R_{\rm planet}}$, $T_{\rm star}$) of the power-law relationship between the effective temperature of their host star and their planetary radii, fitted using our two-level, two-parameter hierarchical model.
     Each blue point represents the mean $T_{\rm eq}$ for a group of hot Jupiters along with the fitted mean value of $\gamma$ (${R_{\rm planet}}$, $T_{\rm star}$) for that group.
     The error bars along the $\gamma$ (${R_{\rm planet}}$, $T_{\rm star}$) axis represent the fitted one-sigma locations, while the error bars along the $T_{\rm eq}$ axis indicate the distribution range of $T_{\rm eq}$ within each group.
     In the middle panel, each blue point along with its error bar shows the mean effective temperature of the host stars for that group and the corresponding distribution range.
     The top panel displays the heating efficiency curves for planets with varying $T_{\rm eq}$ as derived by \citet{Thorngren2018} and \citet{Sarkis2021}. 
              }
         \label{fig5}
   \end{figure*}
   
The three groups of hot Jupiters discussed in section \ref{spectral_teq} represent a coarse grid of planetary  $T_{\rm eq}$.
The three fitted values of $\gamma$ (${R_{\rm planet}}$, $T_{\rm star}$) indicate that the strength of the power-law relationship between planetary radius and stellar effective temperature changes according to  the subpopulation-level $T_{\rm eq}$ of hot Jupiters within a group.
To more precisely assess how the dependence on stellar effective temperature varies with the subpopulation-level $T_{\rm eq}$ for a group of hot Jupiters, we conduct a larger set of similar fittings using a finer grid.
We continuously and overlappingly bin the hot Jupiters based on the amount of normalized flux they have received, with bins covering the ranges of 1-11, 2-12, 3-13, ..., 31-41, and 32-50, respectively.
For all 32 groups, we apply the two-level, two-parameter hierarchical power-law relationship to fit the correlation between stellar effective temperature and planetary radius.
In the fitting results, we focus on the key parameter $\gamma$ (${R_{\rm planet}}$, $T_{\rm star}$), which reflects the strength of the power-law relationship.

Figure \ref{fig5} plots the $\gamma$ (${R_{\rm planet}}$, $T_{\rm star}$) values for all 32 groups. 
As in section \ref{spectral_teq}, we have also converted the incoming stellar flux into the $T_{\textrm{eq}}$ of hot Jupiters, which is represented on the x-axis of the figure, to provide a clearer physical interpretation.
The values of the power-law index $\gamma$ (${R_{\rm planet}}$, $T_{\rm star}$), which quantifies the ralationship between the radius of hot Jupiters and the stellar effective temperature, exhibit a single-peak distribution when plotted against the mean values of $T_{\textrm{eq}}$ of a group of hot Jupiters.
Specifically, it initially increase with rising $T_{\textrm{eq}}$, reaches a peak at approximately 1800 K, and then subsequently decreases. 
The top panel of Figure \ref{fig5} displays the single-peak curves of heating efficiency as derived by \citet{Thorngren2018} and \citet{Sarkis2021}.
Furthermore, the shape of the single-peak distribution of  $\gamma$ (${R_{\rm planet}}$, $T_{\rm star}$) and the position of the peak at $T_{\textrm{eq}}$  show a resemblance to the distribution of the radius anomaly of hot Jupiters as a function of $T_{\textrm{eq}}$ (see Figure 2 in \citet{Laughlin2011}).
The magnitude of the heating efficiency, the radius anomaly, and the power-law index $\gamma$ (${R_{\rm planet}}$, $T_{\rm star}$) are not directly comparable since they represent different physical quantities. 
However, because all these quantities are associated with the heating process of hot Jupiters due to stellar radiation, it is reasonable to compare the characteristics of these curves.
The single-peak distribution of $\gamma$ (${R_{\rm planet}}$, $T_{\rm star}$) with respect to $T_{\textrm{eq}}$ shows a trend that is similar to the single-peak distribution of heating efficiency in relation to $T_{\textrm{eq}}$, though there is a minor discrepancy in the position of the peak.
To quantitatively compare our single-peak distribution of $\gamma$ (${R_{\rm planet}}$, $T_{\rm star}$) with the unimodal distributions of heating efficiency provided by \citet{Sarkis2021} and \citet{Thorngren2018}, we selected and normalized the regions between 800 and 2425 Kelvin from these three distributions,  and then calculated the locations of the 25th, 50th and 75th percentiles for each.
For the heating efficiency distributions, the three quantiles correspond to $T_{\textrm{eq}}$ values of $\sim$ 1485, 1786 and 2062 Kelvin in the results obtained by \citet{Sarkis2021}, while in the findings from \citet{Thorngren2018}, the quantiles are around 1404, 1614 and 1855 Kelvin.
For our single-peak distribution of $\gamma$ (${R_{\rm planet}}$, $T_{\rm star}$), the three quantiles are located at approximately 1259, 1615 and 1821 Kelvin.

At first glance, the single-peak distribution of $\gamma$ (${R_{\rm planet}}$, $T_{\rm star}$), derived from the combination of planetary $T_{\rm eq}$ and stellar effective temperature, appears to replicate the single-peak shape of the distributions of the heating efficiency parameter obtained from planetary thermal evolution models.
However, the following section will show that its interpretation may be misleading due to the influence of several interdependent physical parameters, supported by a more comprehensive analysis.

\section{Single-Peak Distribution: Reanalysis}
\label{sec:analyse}


   \begin{figure*}
   \centering
   \includegraphics[width=4.5in]{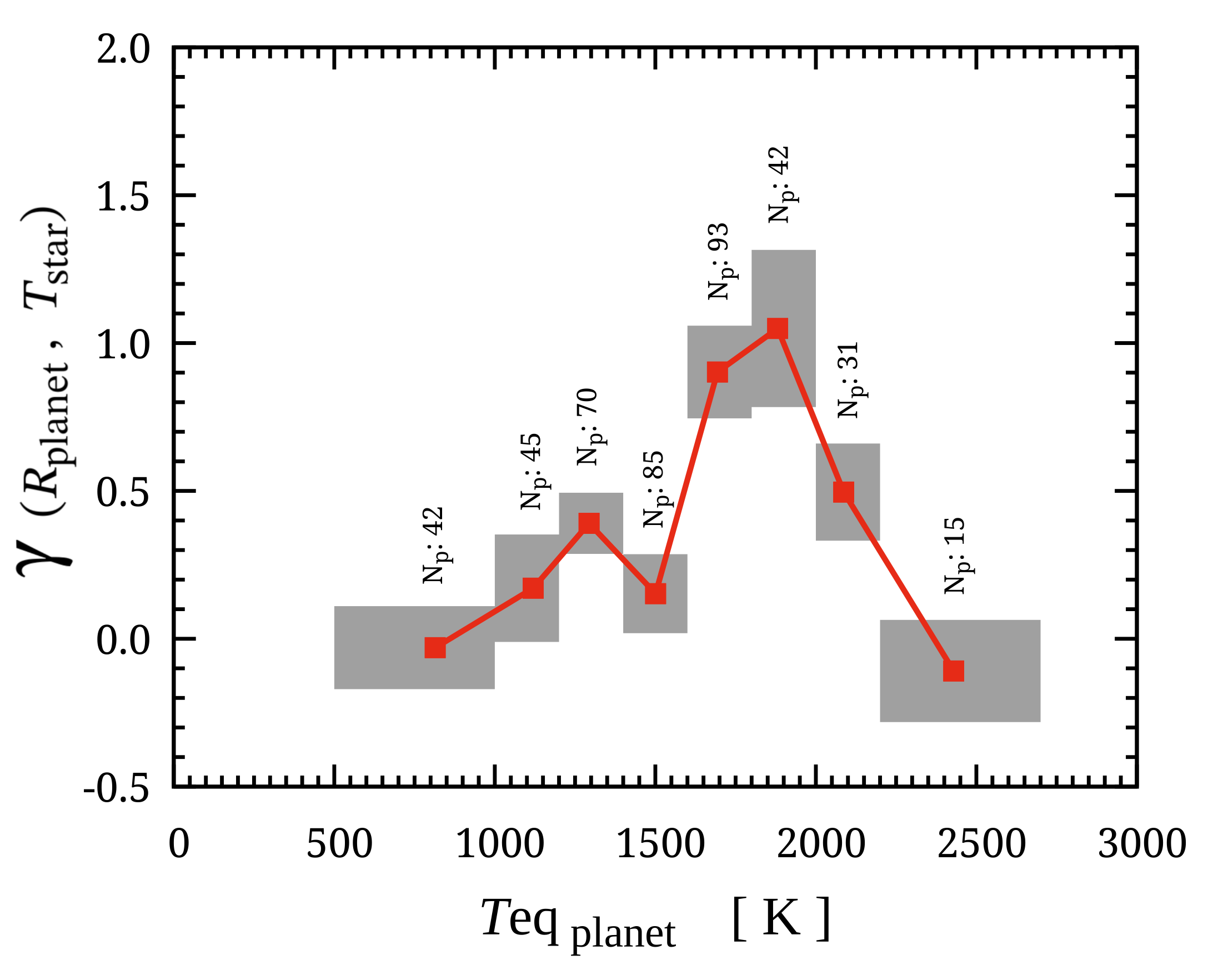}
      \caption{
      The fitted values of $\gamma$ (${R_{\rm planet}}$, $T_{\rm star}$) for eight groups of hot Jupiters created using a separate binning approach.
      The red dots represent the mean values of $T_{\rm eq}$ along with the fitted mean values of $\gamma$ (${R_{\rm planet}}$, $T_{\rm star}$) for each group. 
      The gray rectangles indicate the distribution ranges of $T_{\rm eq}$ for each group of hot Jupiters in the horizontal direction, while in the vertical direction, they reflect the fitted one-sigma ranges of $\gamma$ (${R_{\rm planet}}$, $T_{\rm star}$).
      The vertical labels denote the number of hot Jupiters in each group.
              }
         \label{fig6}
   \end{figure*}

In this section, we re-evaluate the relationship presented in Section \ref{single-peak1} by employing a rigorous binning strategy and incorporating the influence of additional physical quantities.
This approach aims to uncover the underlying factors contributing to the observed single-peak distribution.

\subsection{Separate binning}
\label{sepbin}

We group the hot Jupiters into separate bins to fit the power-law index $\gamma$ (${R_{\rm planet}}$, $T_{\rm star}$) using the hierarchical relationship.
Given the limited total number of confirmed hot Jupiters, we establish a total of 8 bins based on the $T_{\rm eq}$ of hot Jupiters to ensure that each bin contains a sufficient number of planets.
The hot Jupiters are divided into eight bins with $T_{\rm eq}$ ranges of 500 to 1000 K, 1000 to 1200 K, 1200 to 1400 K, 1400 to 1600 K, 1600 to 1800 K, 1800 to 2000 K, 2000 to 2200 K, and 2200 to 2700 K, respectively.
The bins in the middle are spaced 200 Kelvin apart, while the bins at the ends are spaced 500 Kelvin apart due to the small number of planets in those range.
Although the separate binning strategy leads to a coarser grid along the $T_{\rm eq}$ axis, it avoids the complications associate with the moving average that can occur when using overlapping bins. 
Additionally, this approach produces bins with fixed centers and widths.

\begin{table*}
\scriptsize
\centering
\caption{
The $\gamma$ (${R_{\rm planet}}$, $T_{\rm star}$) values (mean $\pm$ standard deviation) obtained from 8 independent sampling procedures within each of the separate bins, along with the corresponding Gelman-Rubin criteria.}
\begin{tabular}{cccccccccc}
\hline
\hline
 $T_{\rm eq}$ (K) & run1 & run2 & run3 &run4 &run5 &run6 &run7 &run8 & $Rc$ value \\
       500 - 1000 & -0.030$\pm0.140$ & 0.058$\pm0.148$ & 0.030$\pm0.169$ & 0.006$\pm0.143$ & 0.038$\pm0.143$ & 0.036$\pm0.139$ & 0.034$\pm0.159$ & 0.030$\pm0.145$ & 1.016 \\
      1000 - 1200 & 0.171$\pm0.182$ & 0.218$\pm0.165$ & 0.182$\pm0.176$ & 0.172$\pm0.148$ & 0.194$\pm0.150$ & 0.210$\pm0.155$ & 0.152$\pm0.117$ & 0.185$\pm0.142$ & 1.009\\
      1200 - 1400 & 0.390$\pm0.103$ & 0.385$\pm0.132$ & 0.402$\pm0.120$ & 0.385$\pm0.110$ & 0.395$\pm0.119$ & 0.383$\pm0.128$ & 0.371$\pm0.122$ & 0.400$\pm0.126$ & 1.004\\
      1400 - 1600 & 0.152$\pm0.134$ & 0.164$\pm0.139$ & 0.172$\pm0.150$ & 0.137$\pm0.135$ & 0.154$\pm0.154$ & 0.115$\pm0.152$ & 0.159$\pm0.125$ & 0.143$\pm0.138$ & 1.008 \\ 
      1600 - 1800 & 0.902$\pm0.157$ & 0.877$\pm0.187$ & 0.893$\pm0.177$ & 0.886$\pm0.128$ & 0.885$\pm0.183$ & 0.886$\pm0.157$ & 0.885$\pm0.179$ & 0.919$\pm0.192$ & 1.003 \\
      1800 - 2000 & 1.049$\pm0.266$ & 1.167$\pm0.252$ & 1.243$\pm0.258$ & 1.104$\pm0.229$ & 1.124$\pm0.212$ & 1.116$\pm0.221$ & 1.120$\pm0.238$ & 1.111$\pm0.241$ & 1.027 \\
      2000 - 2200 & 0.496$\pm0.164$ & 0.476$\pm0.187$ & 0.414$\pm0.205$ & 0.447$\pm0.196$ & 0.486$\pm0.196$ & 0.448$\pm0.170$ & 0.464$\pm0.197$ & 0.467$\pm0.193$ & 1.009\\
      2200 - 2700 & -0.109$\pm0.173$ & -0.108$\pm0.184$ & -0.075$\pm0.209$ & -0.100$\pm0.189$ & -0.143$\pm0.169$ & -0.128$\pm0.172$ & -0.108$\pm0.172$ & -0.096$\pm0.199$ & 1.006 \\
\hline
\end{tabular}
\label{tab:RC}
\end{table*}
\normalsize

   \begin{figure*}
   \centering
   \includegraphics[width=7.0in]{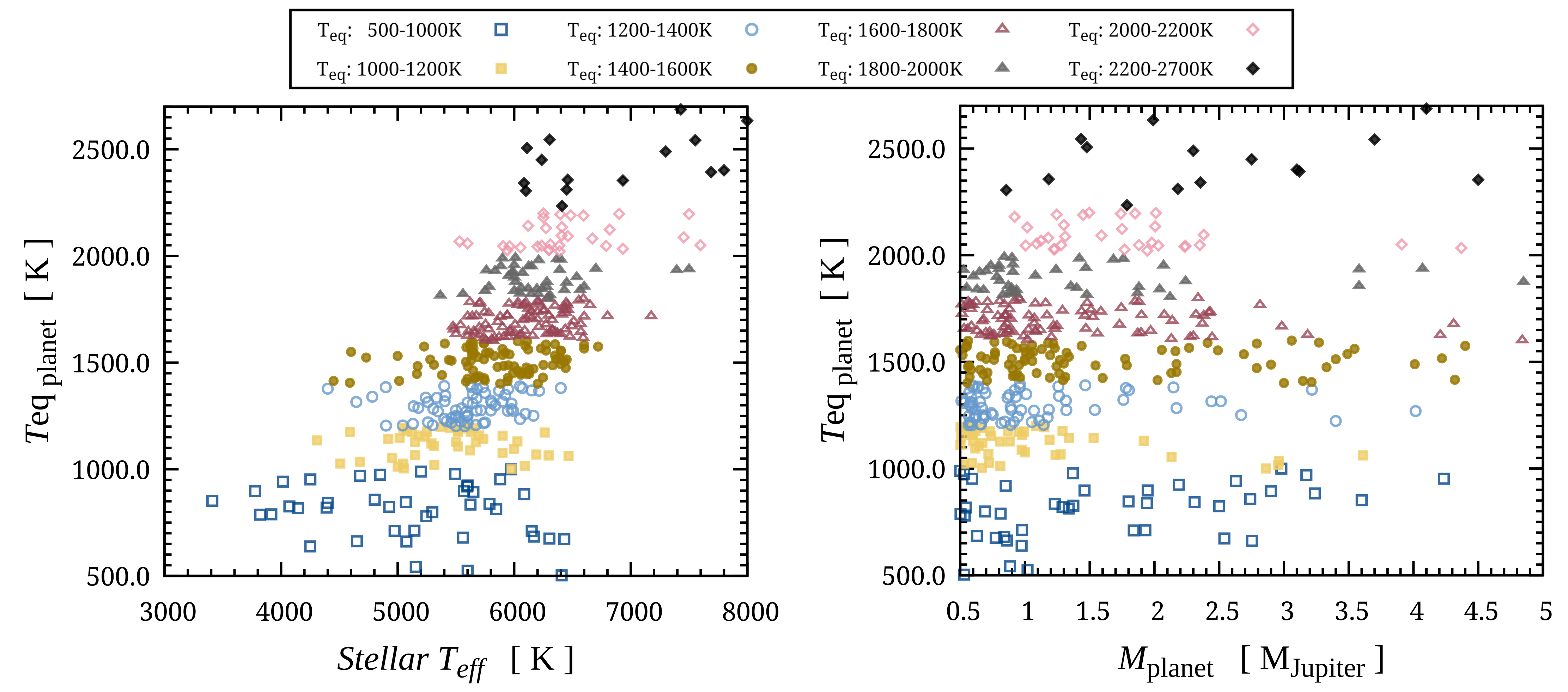}
      \caption{
      The left panel displays the $T_{\rm eq}$ of the hot Jupiters in the 8 separate bins plotted against the effective temperature of their host stars.
The right panel shows the distribution of planetary masses for hot Jupiters across each of the 8 bins, indicating that the masses of hot Jupiters with ultra-high $T_{\rm eq}$ are significantly larger.
              }
         \label{fig7}
   \end{figure*}

To ensure that our sampling yielded convergent results, we conducted 8 independent sampling processes using different initial random seeds for fitting each bin of hot Jupiters.
For each of the 8 bins, we gathered the means and standard deviations of $\gamma$ (${R_{\rm planet}}$, $T_{\rm star}$) from the 8 independent runs.
Subsequently, the Gelman-Rubin criteria  \citep{Gelman1992} were utilized to evaluate the convergence of the Markov chains in each bin.
The results are presented in Table \ref{tab:RC}, which demonstrates that all our MCMC samplings produced convergent results, as all calculated Gelman–Rubin criteria are bellow 1.03.
This indicates that the MCMC results satisfy the convergence criteria for rigorous diagnostics \citep{Gelman1992,Brooks1998}.

Figure \ref{fig6} displays the fitted values of $\gamma$ (${R_{\rm planet}}$, $T_{\rm star}$) for these 8 groups.
Similar to Figure \ref{fig5}, a single-peak distribution is observed, with the highest $\gamma$ (${R_{\rm planet}}$, $T_{\rm star}$)  value recorded for hot Jupiters with $T_{\rm eq}$ ranging from 1800 to 2000 K.
Figure \ref{fig6} also plots the one-sigma ranges of the fitted $\gamma$ (${R_{\rm planet}}$, $T_{\rm star}$) for each group. 
The results show that the hot Jupiters with $T_{\rm eq}$ ranging from 1600 to 2000 K exhibit significantly higher $\gamma$ (${R_{\rm planet}}$, $T_{\rm star}$) values compared to other groups, even when accounting for the one-sigma fluctuations.

\subsection{The rising part: incident flux}

\begin{table*}
\scriptsize
\centering
\caption{
The fitted means and standard deviations of three important $\gamma$ in the 8 bins.
}
\begin{tabular}{ccccccccc}
\hline
\hline
  & 500-1000K & 1000-1200K & 1200-1400K & 1400-1600K & 1600-1800K & 1800-2000K & 2000-2200K & 2200-2700K \\
\hline
$\gamma$ (${T_{\rm eq}}$, ${T_{\rm star}}$) & -0.086$\pm0.120$ & -0.053$\pm0.061$ &  -0.029$\pm0.054$ & 0.068$\pm0.043$ & 0.121$\pm0.053$ & 0.093$\pm0.055$ &  0.110$\pm0.064$ & 0.189$\pm0.092$  \\
$\gamma$ (${R_{\rm planet}}$, ${M_{\rm planet}}$) & -0.011$\pm0.033$ & 0.019$\pm0.026$ & -0.017$\pm0.018$ & -0.009$\pm0.018$ & -0.129$\pm0.018$ & -0.094$\pm0.035$ & -0.024$\pm0.044$ & -0.135$\pm0.042$ \\
$\gamma$ (${R_{\rm planet}}$, ${T_{\rm star}}$) & -0.030$\pm0.140$ & 0.171$\pm0.182$ & 0.390$\pm0.103$ & 0.152$\pm0.134$ & 0.902$\pm0.157$ & 1.049$\pm0.266$ & 0.496$\pm0.164$ & -0.109$\pm0.173$ \\
\hline
\end{tabular}
\label{tab:2power}
\end{table*}
\normalsize

Section \ref{single-peak1} and \ref{sepbin} have shown that a similar single-peak distribution of the planetary radius expansion efficiency along $T_{\rm eq}$ can be observed due to the influence of stellar effective temperature.
However, it is important to verify the validity of this finding, as the radius of a hot Jupiter is influenced by numerous factors.
An implicit interdependent parameter that may be overlooked is the amount of incident flux received by a planet, as the strength of stellar flux is correlated with stellar effective temperature.
Consequently, we investigate whether the single-peak distribution we identified is influenced by an implicit dependence on incident flux.
Specifically, for the 8 subpopulations of hot Jupiters in Figure \ref{fig6} across different $T_{\rm eq}$ bins, how do the relationships, at the subpopulation level, between incident flux and stellar effective temperature vary?

The left panel of Figure \ref{fig7} presents the $T_{\rm eq}$ distribution of the 8 subpopulations of hot Jupiters  plotted against the effective temperature of their host stars.
We employed the same two-level, two-parameter hierarchical model to fit the power-law relationship between planetary $T_{\rm eq}$ and stellar effective temperature for these 8 bins of hot Jupiters, and collect the means and standard deviations of the power-law index $\gamma$ (${T_{\rm eq}}$, ${T_{\rm star}}$).
Table \ref{tab:2power} presents the means and standard deviations of $\gamma$ (${T_{\rm eq}}$, ${T_{\rm star}}$) obtained from the 8 bins. 
The means for these 8 bins, listed from the lowest $T_{\rm eq}$ to the highest, are 
-0.086$\pm0.120$, -0.053$\pm0.061$,  -0.029$\pm0.054$, 0.068$\pm0.043$, 0.121$\pm0.053$, 0.093$\pm0.055$,  0.110$\pm0.064$,  and 0.189$\pm0.092$, respectively. 
It is noteworthy that these values exhibit a monotonically increasing trend across  the entire range of $T_{\rm eq}$, reaching the highest value in the 2200-2700K group.
Since different $T_{\rm eq}$ correspond to varying amounts of incident flux, 
our result indicates that the relationships between incident flux and stellar effective temperature  change across the 8 subpopulations of hot Jupiters in different $T_{\rm eq}$ bins.
In general, for planet subpopulations in higher $T_{\rm eq}$  bins, the incident flux they receive becomes increasing dependent on the effective temperature of their host stars.
As a consequence, the planetary radii will also become more influenced by the stellar temperature.
This implies that the rising part of the single-peak distribution observed in Sections \ref{single-peak1} and \ref{sepbin} is primarily driven by the correlation between incident flux and stellar effective temperature.

This conclusion is easier to gasp from a physical standpoint. 
First, the intensity of the incoming flux received by a hot Jupiter is correlated with the effective temperature of its host star. 
Second, a higher planetary $T_{\rm eq}$ typically indicates that the planet is closer to its host star, making its structure  more sensitive to variations in the incoming stellar flux.
The combination of these two factors leads to the observed rising part of the single-peak distribution of $T_{\rm eq}$, which spans from approximately 1500 K to 1800 K, where the radius expansion of hot Jupiters is more efficient.
But why is the radius expansion efficiency of hot Jupiters observed to decrease at even higher $T_{\rm eq}$ above 2000 K?

\subsection{The falling part: planetary mass}

The reason for employing a two-parameter model in our fit is due to the finite number of available hot Jupiters available. 
We do not wish to further subdivide our sample by introducing additional physical dependencies, as our goal is to perform a statistical analysis based on a population with the largest possible number of cases.
However, when we split the hot Jupiters into smaller bins in Sections \ref{single-peak1} and \ref{sepbin}, the number of effective samples is further reduced.
Therefore, we must be cautious not to introduce additional bias by relying  on an even smaller set of samples.
Another important physical quantity of hot Jupiters is the planetary mass.
While Section \ref{pmass} demonstrates a weak correlation between the population-wide mean radius and planetary mass, this correlation may not be applicable to smaller sample sizes.
The right panel of Figure \ref{fig7} shows the distribution of planetary mass for hot Jupiters across the 8 separate bins. 
It indicates that for hot Jupiters with $T_{\rm eq}$ above 2000 K, the distribution of planetary mass does not follow the same trend as that observed for hot Jupiters with lower $T_{\rm eq}$; specifically, there is a noticeable absence of planets with lower masses.
Indeed, when we calculate the mean planetary masses for hot Jupiters in the 8 bins, the values obtained are 1.628, 1.073,  1.091, 1.590, 1.364, 1.411, 1.771, and 2.456 (${\rm M}_{\rm Jupiter}$), respectively.
The large planetary masses at ultra-high $T_{\rm eq}$ can dampen the radius expansion of hot Juipiters, as planetary mass is a crucial factor in determining the extent of a planet's radius enlargement at a particular irradiation flux \citep{Sestovic2018}.

To investigate how the mass distribution of the limited number of hot Jupiters in the 8 bins may influence their planetary radius relationship, we applied the same two-level, two-parameter hierarchical Bayesian model to fit the mass-radius relationship for the 8 separate bins.
We employed the same approach as in Section \ref{pmass}, albeit with smaller sample sizes.
Table \ref{tab:2power} presents the means and standard deviations of the fitted power-law index $\gamma$ (${R_{\rm planet}}$, ${M_{\rm planet}}$).
The means of the fitted $\gamma$ values for the 8 bins, ordered from the lowest $T_{\rm eq}$ to the highest, are 
-0.011$\pm0.033$, 0.019$\pm0.026$, -0.017$\pm0.018$, -0.009$\pm0.018$, -0.129$\pm0.018$, -0.094$\pm0.035$, -0.024$\pm0.044$, and -0.135$\pm0.042$, respectively. 
Clearly, as the number of planets in each bin decreases, the correlation between planetary mass and radius becomes stronger. 
This is especially pronounced for the bin with $T_{\rm eq}$ in the range of 2200-2700 K, where the fitted $\gamma$ is -0.135, indicating a significant reduction in radius at larger planetary masses.
Furthermore, the masses of the hot Jupiters in the two bins with planetary $T_{\rm eq}$ greater than 2000 K are significantly larger compared to those in the other bins.
These factors, when combined, lead to the substantial decrease in radius expansion efficiency observed in Figures \ref{fig5} and \ref{fig6}.

The analysis in this section elucidates the reasons for the decrease in the radius expansion efficiency of hot Jupiters at high $T_{\rm eq}$, based on the mass distribution of the currently detected hot Jupiters across different $T_{\rm eq}$ bins.
However, this does imply that hot Jupiters are inherently more massive at higher $T_{\rm eq}$, rather than being solely the result of an observational bias.
Because hotter planets are predominantly found around hotter stars with effective temperatures exceeding 6000 Kelvin, and it is known that radial velocity measurements become increasingly difficult as stars exceed the Kraft break at approximately 6250 Kelvin \citep{Kraft1967}.
%

\section{Discussion and Summary}
\label{sec:summary}

The primary aim of this work is to uncover the underlying cause of the single-peak distribution of heating efficiency of hot Jupiters, as inferred by previous studies.
We fit the relationships between different pairs of observed physical quantities using two-layer, two-parameter hierarchical Bayesian models.
Our analysis is solely based on fitting of observed data and does not depend on any theoretical model. 
We find that a parameter characterizing the planetary radius expansion efficiency, specifically the power-law index $\gamma$  (${R_{\rm planet}}$, ${T_{\rm star}}$) of the radius relationship, exhibits a single-peak shape along the planetary $T_{\rm eq}$.
This feature resembles previously derived heating efficiency parameters found in complex inference models that incorporate planetary thermal evolution \citep{Thorngren2018,Sarkis2021}. 
Through a comprehensive analysis, we find that this single-peak distribution can be explained by straightforward physical processes. 
The rising portion of the single-peak distribution can be attributed to the increase in incident stellar flux, while the the falling part is associated with  the increase in gravitational binding energy related to larger planetary masses.

An important implication of this work is the necessity for caution when making statistical inferences about exoplanet populations, since many planetary physical quantities are interdependent, and there exist subpopulations of exoplanets that exhibit distinct distributional properties.
For instance, our initial finding indicates that the single-peak distribution of the radius expansion efficiency of hot Jupiters can be explained by the strength of the correlation with the effective temperature of their host stars.
However, further analysis reveals a gradual strengthening of the correlation between incident stellar flux and stellar effective temperature as the planetary $T_{\rm eq}$ increases.
Additionlly, the planetary masses of the subpopulation of hot Jupiters with $T_{\rm eq}$ above 2000 K are considerably larger.
Because the incident flux and planetary masses are more fundamental quantities that can influence the structure of a planet, they are likely to be the primary factors contributing to the observed single-peak distribution of heating efficiency.

Although the effects of incident flux and planetary mass offer a more fundamental physical explanation for the radius expansion efficiency of hot Jupiters, it is still possible that stellar effective temperature may also play a role, as indicated by the distribution of $\gamma$ (${R_{\rm planet}}$, ${T_{\rm star}}$)  along the planetary $T_{\rm eq}$ in Figures \ref{fig6} and \ref{fig7}.
On the one hand, different stellar effective temperatures correspond to different stellar spectral types, which can affect the specific details of the radiative transfer process occurring in planetary atmospheres.
Theoretical studies indicate that the dissociation and recombination of hydrogen  in ultra-hot Jupiters influence the atmospheric structure and the heat transport efficiency from the dayside to the nightside \citep{Bell2018,Komacek2018,Tan2019,Roth2021}.
Consequently, the atmospheric structure and dynamics of ultra-hot Jupiters are highly sensitive to both the planetary $T_{\textrm{eq}}$ and the type of host star \citep{Tan2024}.
On the other hand, observations have shown that the spectra of ultra-hot Jupiters  appear featureless throughout the 1.1-1.7 $\mu m$ region \citep{Arcangeli2018,Kreidberg2018,Mansfield2018}.
This is because the majority of molecules are thermally dissociated, and alkaline elements are ionized within the day-side photospheres of ultra-hot Jupiters, resulting in continuum opacity due to dissociated hydrogen \citep{Bell2017,Lothringer2018,Parmentier2018}. 
From this perspective, the radiative transfer process in the atmospheres of ultra-hot Jupiters will be less sensitive to differences in host star types.

It is also noteworthy that subpopulations of hot Jupiters exhibit distinct distribution characteristics. 
For instance, the right panel of Figure \ref{fig7} shows that the mass distribution of hot Jupiters in the subpopulations with $T_{\rm eq}$  above 2000 K differ from that of subpopulations with lower $T_{\rm eq}$.
The masses of the detected hot Jupiters with ultra-high $T_{\rm eq}$ are significantly larger.
First, detecting planets through the radial velocity method is extremely challenging around hot stars, which introduces observational biases.
Second, gaseous planets with high $T_{\rm eq}$ orbiting hotter stars may have experienced significant atmospheric escape processes, which can subsequently influence their final physical characteristics \citep{Baraffe2004}.
Ignoring the differences in the distributional characteristics of the subpopulations of hot Jupiters in statistical inference could result in erroneous conclusions.

Since the results presented in this work are derived exclusively from statistical analyses of observational data pertaining to hot Jupiter populations, our findings can serve as a useful reference for future theoretical studies on hot Jupiters.

\vspace{0.9cm}
We thank the anonymous referees for their constructive comments,  that have significantly improved this paper.
This work is supported by the National Natural Science Foundation of China (Nos. 11973094, 12103003), the Youth Innovation Promotion Association CAS (No. 2020319), the incubation program for recruited talents (No. 2023GFXK153), and the doctoral start-up funds from Anhui Normal University. 
S.J., Y.C and Z.G. acknowledge support from Zijin exploration program provided by the High School Affiliated to Nanjing Normal University and the Nanjing Branch of the Chinese Academy of Sciences.

\software{\texttt{Nii-C} \citep{Jin2024}, \texttt{gnuplot} \citep{Williams2022}}

\bibliographystyle{aasjournal}

\end{document}